\documentstyle[epsfig,longtable]{aipproc}

\begin{document}
\title{Rotational Instabilities\\ in Post--Collapse Stellar Cores}

\author{J. David Brown}
\address{Department of Physics, North Carolina State University,
Raleigh, NC 27695-8202}

\maketitle

\begin{abstract}
A core--collapse supernova might produce large amplitude gravitational waves if, through 
the collapse process, the inner core can aquire enough rotational energy to become
dynamically unstable. In this report I present the results of 3-D numerical simulations 
of core collapse supernovae. These simulations indicate that for some initial conditions 
the post--collapse inner core is indeed unstable. However, for the cases considered, the 
instability does not produce a large gravitational--wave signal.
\end{abstract}

\section*{Introduction}

Core--collapse supernovae are potentially rich sources of gravitational waves. When a 
rotating stellar core exhausts its nuclear fuel the matter in the polar regions 
collapses more rapidly than the matter in the equatorial plane, since the latter must 
fight harder against centrifugal forces as it spirals inward. The changing  
oblateness of the core during this infall phase, and the subsequent changes 
due to core bounce, will generate gravitational waves (see, for example, 
Refs.~\cite{dbrown:YandK,dbrown:ZandM}). After core bounce convective instabilities 
will cause the 
hot neutron star to boil \cite{dbrown:bethe}, and the resulting convective motions 
will give rise to a gravitational--wave signal \cite{dbrown:MandJ}. 
To a lesser extent, the same boiling process and release of gravitational waves 
can occur in the neutrino--heated material 
interior to the shock wave \cite{dbrown:MandJ}. The supernova explosion itself may 
have a preferred direction, as evidenced by the large kick 
velocities of many neutron stars \cite{dbrown:tauris}. Such an asymmetry will generate 
a gravitational--wave signal with memory \cite{dbrown:BandH}. Perhaps the most interesting 
possibility, and the one discussed in this paper, is that the stellar core will spin 
up as it collapses and produce a very 
rapidly rotating neutron star. The neutron star might be subject to dynamical 
instabilities that act to deform, or even fragment the star, and in the process 
produce large amplitude gravitational waves. 

In this paper I present the results of numerical 
simulations aimed at determining 
the types of initial conditions for a pre--collapse stellar core that lead to a 
dynamically unstable post--collapse inner core. The only previous investigations along 
these lines is found in the work of Rampp, M{\" u}ller, and Ruffert \cite{dbrown:RMR}.

Analytical and numerical work on 
rapidly rotating fluid stars with Maclaurin--like rotational laws has shown that 
dynamical instabilities, in particular the 
$m=2$ bar--mode instability, will grow when the stability parameter $T/|W|$ (the 
ratio of rotational kinetic 
energy to gravitational potential energy) exceeds about $0.27$ \cite{dbrown:tassoul}. 
For certain angular velocity profiles, a dynamical $m=1$ instability can grow for $T/|W|$ as low 
as $\sim 0.14$ \cite{dbrown:modeone}. 
A back--of--the--envelope calculation
frequently quoted in the literature suggests that the stability parameter will scale as 
$T/|W| \sim 1/R$ during collapse, where $R$ is the radius of the core. For a $10$ solar 
mass star whose core collapses completely to neutron star size, this implies a factor $\sim 100$ 
increase in $T/|W|$. Based on this argument, one would expect that even the most slowly 
rotating cores will be dynamically unstable after collapse. 

The factor $\sim 100$ increase in $T/|W|$ is 
overly optimistic for two reasons. First, centrifugal forces can halt the collapse at 
subnuclear density, producing a bloated inner core with $T/|W|$ below the threshold for dynamical 
instability.  Second, the core actually ``implodes", rather than collapses, and 
some of the stellar core's matter and angular momentum remain outside the inner core. Since 
only a percentage of the core's angular momentum is drawn into the inner core, the 
increase in rotational kinetic energy is less than one would predict by assuming a 
more complete collapse. On the other hand, the simulations described here provide evidence 
that the entire core need not have $T/|W| \gtrsim 0.27$ for the bar mode to grow on a relatively 
short timescale. It might be possible for the bar instability to grown 
as a secular process on a timescale of a few rotation periods, 
due to coupling between the inner and outer core regions. 

Two of the most rapidly rotating pre--collapse models considered here lead to dynamically 
unstable post--collapse inner cores. These models are unstable to the $m=2$ bar mode, but the 
resulting bars have insufficient density and spatial extent to generate large gravitational--wave 
signals. In both cases, the contribution to the gravitational--wave signal from the bar 
deformation is smaller in amplitude, by a factor of $\sim 2$--$5$, than the purely axisymmetric 
signal generated by the core's changing oblateness during collapse and bounce. Overall, 
for sources in the Virgo cluster, these 
signals are too weak to be detected by laser interferometers operating in broadband mode. For 
galactic sources, these signals should be detectable. 

\section*{Numerical Code and Initial Models}

The numerical code used for these investigations was written in collaboration with John Blondin. 
It uses Newtonian hydrodynamics and Newtonian gravity, ignores neutrino heating and cooling, and 
uses the relatively simple analytical equation of state discussed by Zwerger and M{\" u}ller 
\cite{dbrown:ZandM,dbrown:RMR}. The gravitational--wave signal is computed in the quadrupole 
approximation.  The hydrodynamical equations are solved with VH-1, written by Blondin, J.~Hawley, 
G.~Lindahl, and E.~Lufkin. VH-1 uses the piecewise parabolic method \cite{dbrown:CandW}. The 
Poisson equation for the gravitational potential is solved with multigrid techniques. 

Our code models the supernova in a minimal way, retaining just enough physics to capture the 
gravitational--wave signal in the leading--order quadrupole approximation. Nevertheless, the 
computation is a challenge due to the discrepancy in length scales involved. While the stellar 
core has a radius of $1000$'s of kilometers, the computational 
grid must have zone sizes of no more than $\sim 1\,{\rm km}$  to support the 
steep density and velocity gradients in the inner core. A uniform 3-D Cartesian grid would require 
$\sim 10^{10}$ zones for these simulations, which is not feasible with current technology. 
One possible solution to this problem is to use a spherical--coordinate grid with non--uniform 
radial spacing, so the zones are smallest in the inner regions of the grid. The difficulty with 
this approach is that the zones become too narrow in the angular 
directions near the coordinate origin, and this drives the Courant--limited timestep to zero. The 
solution we have adopted is a nested grid scheme, in which the computational domain is covered 
by a sequence of Cartesian grids with 
increasing resolution and decreasing size. In this way only the inner--most region is covered with 
high resolution. This approach was also used by Rampp, M{\" u}ller, 
and Ruffert \cite{dbrown:RMR}. The simulations described here use $7$ grids, each with $64^3$ zones. 

In addition to the 3-D code just described, I also work with a 2-D code that assumes 
axisymmetry. The 2-D code uses a system of nested Cartesian grids in the $r$--$z$ plane to achieve 
the necessary resolution of the inner core. The 2-D simulations reported here use $6$ 
grids, each with $128^2$ zones. 

In this paper I consider a sequence of initial models of the pre--collapse stellar core. The 
goal is to determine which of these models lead to dynamically unstable post--collapse inner cores. 
Each initial model is a rotating, equilibrium polytrope 
with $\Gamma = 4/3$ and central density $\rho_c = 10^{10}\,{\rm g}/{\rm cm}^3$. The rotation 
laws for the sequence are given by 
\begin{equation}
   \Omega(\varpi) = \Omega_0 e^{-(\varpi/1500\,{\rm km})^2} \ ,\qquad 
   \Omega_0 = 8,\ 12,\ 16,\ 20,\ 24\,{\rm rad}/{\rm s},
\end{equation}
where $\Omega$ is the angular velocity and $\varpi$ is the distance from the rotation axis. 
The Gaussian form for $\Omega(\varpi)$ was motivated in part by the results of Heger, Langer, and 
Woosley \cite{dbrown:HLW}. Their simulations of stellar evolution with rotation yield pre--collapse 
cores with relatively broad, Gaussian--like angular velocity profiles, similar in shape to the 
profile (1). Note, however, that even their 
most rapidly rotating model has an overall scale of $\Omega_0 \approx 10$, somewhat smaller 
than most of the models (1).

In their previous work Rampp, M{\" u}ller, and Ruffert \cite{dbrown:RMR} used an initial data 
set with angular velocity profile $\Omega(\varpi) = \Omega_0/[1 + (\varpi/100\,{\rm km})^2]$ 
and $\Omega_0 \approx 140\,{\rm rad}/{\rm s}$. 
The resulting model is highly differentially rotating. The angular velocity drops from 
its central value of about $140\,{\rm rad}/{\rm s}$ to less than $4\,{\rm rad}/{\rm s}$ at 
a distance of $R_{\rm eq}/2$, where $R_{\rm eq} \approx 1260\,{\rm km}$ is the equatorial 
radius. By contrast, the angular velocity of the 
most rapidly rotating model (1) drops from $24\,{\rm rad}/{\rm s}$ to about $13\,{\rm rad}/{\rm s}$
at $R_{\rm eq}/2$, where $R_{\rm eq} \approx 2500\,{\rm km}$. 

The equation of state \cite{dbrown:ZandM,dbrown:RMR} contains a polytropic part whose stiffness 
depends on whether the matter is below or above nuclear density, and a thermal part that 
models the thermal pressure of matter heated by shock waves. Collapse of the initial models 
is induced by choosing the polytropic index in the sub--nuclear density regime to be $\Gamma = 1.28$, 
significantly below the value of $4/3$ required for the model to remain in equilibrium. Random 
density perturbations at the $1\%$ level were imposed at the beginning of the 3-D simulations. 

\section*{Results}

The stability parameter is plotted as a function of time in 
Figure 1 for the sequence of models (1). These simulations 
were run with the 2-D code. 
Observe that $T/|W|$ increases by a factor 
of $2$ or less as a result of core collapse. 
As shown in Figure 2, for the initial models 
with angular velocity $\Omega_0 = 16$, $20$ and $24\,{\rm rad}/{\rm s}$, 
the core experiences a centrifugal bounce and never 
reaches nuclear density $\rho_{\rm nuc} = 2.0\times 10^{14}\,{\rm g}/{\rm cm}^3$. 
For the model with $\Omega_0 = 12\,{\rm rad}/{\rm s}$, the inner core reaches 
nuclear density at core bounce then relaxes to a central value slightly below nuclear density. 
Only the most slowly rotating initial data, with $\Omega_0 = 8\,{\rm rad}/{\rm s}$, 
forms a stiff inner core with density greater than $\rho_{\rm nuc}$. These results show that 
centrifugal forces can severely inhibit core collapse, and prevent the stability parameter 
from experiencing the kind of growth suggested by the back--of--the--envelope arguments  
discussed in the introduction.  
\begin{figure}[b] 
\begingroup%
  \makeatletter%
  \newcommand{\GNUPLOTspecial}{%
    \@sanitize\catcode`\%=14\relax\special}%
  \setlength{\unitlength}{0.1bp}%
\begin{picture}(3600,2160)(0,0)%
\special{psfile=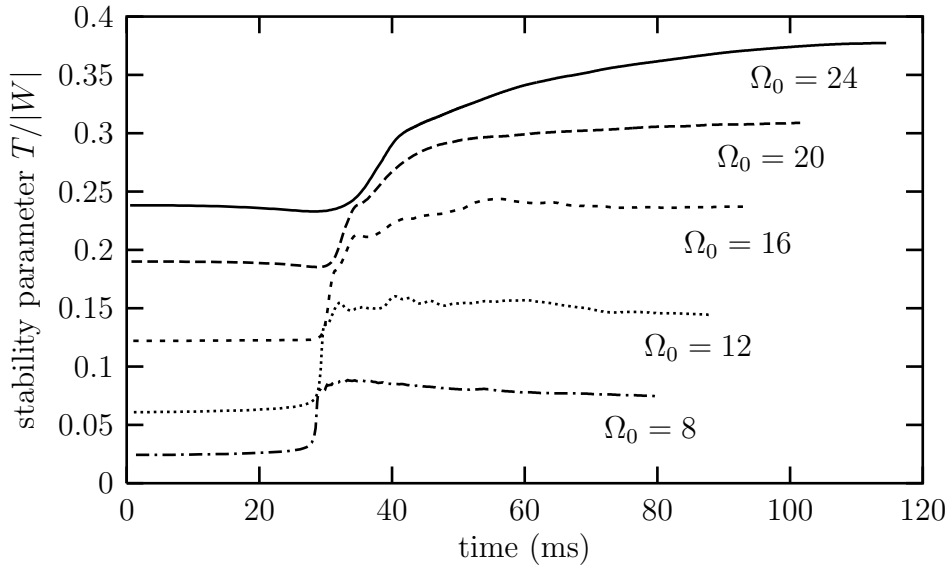 llx=0 lly=0 urx=720 ury=504 rwi=7200}
\put(2250,500){\makebox(0,0)[l]{$\Omega_0 = 8$}}%
\put(2400,810){\makebox(0,0)[l]{$\Omega_0 = 12$}}%
\put(2550,1200){\makebox(0,0)[l]{$\Omega_0 = 16$}}%
\put(2675,1525){\makebox(0,0)[l]{$\Omega_0 = 20$}}%
\put(2800,1820){\makebox(0,0)[l]{$\Omega_0 = 24$}}%
\put(1950,50){\makebox(0,0){time (ms)}}%
\put(100,1180){%
\special{ps: gsave currentpoint currentpoint translate
270 rotate neg exch neg exch translate}%
\makebox(0,0)[b]{\shortstack{stability parameter $T/|W|$}}%
\special{ps: currentpoint grestore moveto}%
}%
\put(3450,200){\makebox(0,0){120}}%
\put(2950,200){\makebox(0,0){100}}%
\put(2450,200){\makebox(0,0){80}}%
\put(1950,200){\makebox(0,0){60}}%
\put(1450,200){\makebox(0,0){40}}%
\put(950,200){\makebox(0,0){20}}%
\put(450,200){\makebox(0,0){0}}%
\put(400,2060){\makebox(0,0)[r]{0.4}}%
\put(400,1840){\makebox(0,0)[r]{0.35}}%
\put(400,1620){\makebox(0,0)[r]{0.3}}%
\put(400,1400){\makebox(0,0)[r]{0.25}}%
\put(400,1180){\makebox(0,0)[r]{0.2}}%
\put(400,960){\makebox(0,0)[r]{0.15}}%
\put(400,740){\makebox(0,0)[r]{0.1}}%
\put(400,520){\makebox(0,0)[r]{0.05}}%
\put(400,300){\makebox(0,0)[r]{0}}%
\end{picture}%
\endgroup
 
\smallskip
\caption{$T/|W|$ vs.~$t$ for the sequence of models (1).}
\end{figure}

\begin{figure}[t] 
\begingroup%
  \makeatletter%
  \newcommand{\GNUPLOTspecial}{%
    \@sanitize\catcode`\%=14\relax\special}%
  \setlength{\unitlength}{0.1bp}%
\begin{picture}(3600,2160)(0,0)%
\special{psfile=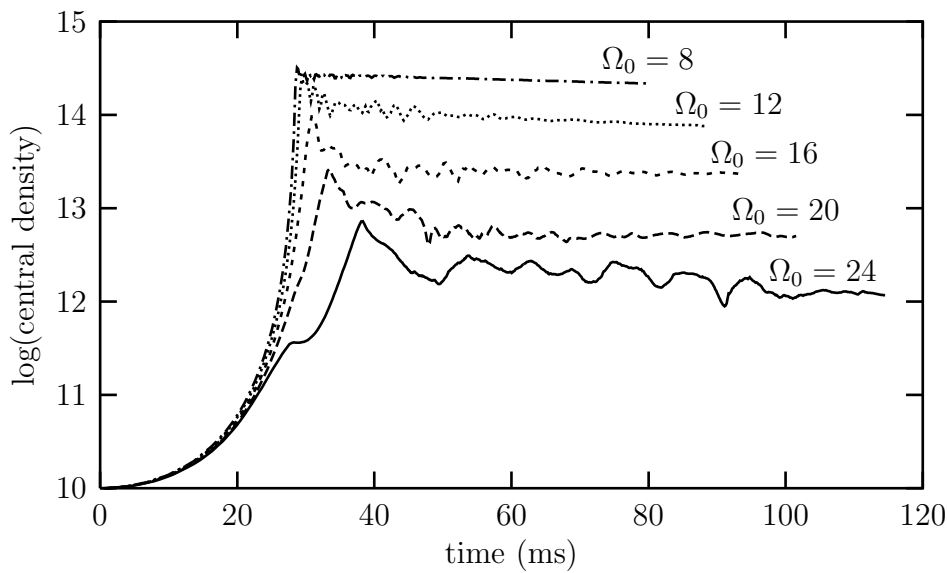 llx=0 lly=0 urx=720 ury=504 rwi=7200}
\put(2240,1915){\makebox(0,0)[l]{$\Omega_0 = 8$}}%
\put(2515,1745){\makebox(0,0)[l]{$\Omega_0 = 12$}}%
\put(2650,1560){\makebox(0,0)[l]{$\Omega_0 = 16$}}%
\put(2730,1345){\makebox(0,0)[l]{$\Omega_0 = 20$}}%
\put(2870,1115){\makebox(0,0)[l]{$\Omega_0 = 24$}}%
\put(1900,50){\makebox(0,0){time (ms)}}%
\put(100,1180){%
\special{ps: gsave currentpoint currentpoint translate
270 rotate neg exch neg exch translate}%
\makebox(0,0)[b]{\shortstack{log(central density)}}%
\special{ps: currentpoint grestore moveto}%
}%
\put(3450,200){\makebox(0,0){120}}%
\put(2933,200){\makebox(0,0){100}}%
\put(2417,200){\makebox(0,0){80}}%
\put(1900,200){\makebox(0,0){60}}%
\put(1383,200){\makebox(0,0){40}}%
\put(867,200){\makebox(0,0){20}}%
\put(350,200){\makebox(0,0){0}}%
\put(300,2060){\makebox(0,0)[r]{15}}%
\put(300,1708){\makebox(0,0)[r]{14}}%
\put(300,1356){\makebox(0,0)[r]{13}}%
\put(300,1004){\makebox(0,0)[r]{12}}%
\put(300,652){\makebox(0,0)[r]{11}}%
\put(300,300){\makebox(0,0)[r]{10}}%
\end{picture}%
\endgroup
 
\smallskip
\caption{$\ln(\rho_c)$ vs.~$t$ for the sequence of models (1).}
\end{figure}
The stability parameter for the two most rapidly rotating 
models exceeds $0.27$ after core bounce. I will use the 
labels $\Omega 20$ and $\Omega 24$ to denote the models with 
$\Omega_0 = 20\,{\rm rad}/{\rm s}$ and $\Omega_0 = 24\,{\rm rad}/{\rm s}$, respectively. 
For these models one might expect the post--bounce inner core, although centrifugally 
hung at sub--nuclear densities,  to be dynamically unstable to growth of the $m=2$ bar mode. 
The inner cores are not likely to be unstable to 
the $m=1$ mode discussed in Reference \cite{dbrown:modeone}, since their density profiles 
are centrally peaked. Note, however, that prior to collapse both of these models exceed 
the nominal threshold $T/|W| \approx 0.14$ for growth of secular instabilities. Thus, these models are 
unrealistic---a realistic stellar core with such a high rate of rotation would lose its axisymmetry 
prior to collapse. With this caveat in mind, I have forged ahead and evolved these models with 
the 3-D code to check for dynamical instabilities in the post--collapse inner cores. 
These simulations show that model $\Omega 20$ remains 
dynamically stable after collapse. Model $\Omega 24$
is unstable after collapse, and its inner core deforms into a bar shape. 

For the model considered by Rampp, M{\" u}ller, and Ruffert (RMR) \cite{dbrown:RMR}, the 
pre--collapse core has an initial value for $T/|W|$ of about $0.04$, 
well below the threshold for growth of secular instabilities. Due to the high degree of 
differential rotation, the material in the outer layers rotates 
relatively slowly. As collapse proceeds, the lack of substantial centrifugal support allows 
the matter in the outer layers to strongly compress the inner core. The result 
is that the stability parameter increases by a larger fraction 
than that obtained with the data sets (1). 
The peak value of $T/|W|$ is about $0.35$, although $T/|W|$ stays above $0.27$ for less 
than $2\,{\rm ms}$. After bounce, the inner core relaxes and $T/|W|$ quickly settles to 
a value of about $0.19$. 

In their 3-D simulations, RMR imposed $10\%$ random and $5\%$ $m=3$ density perturbations 
at a time of $2.5\,{\rm ms}$ before core bounce. The inner core showed $m=2$, $3$, and $4$ 
asymmetries after bounce, but no significant enhancement in the gravitational--wave 
signal. Their simulations were halted at about $45\,{\rm ms}$, approximately $15\,{\rm ms}$ 
after core bounce. The following 
question naturally arises: are the post--bounce asymmetries seen in the RMR simulation merely 
a transient effect caused by the asymmetrical bounce?  I have 
carried out a 3-D simulation with the RMR initial data using $1\%$ random density perturbations 
imposed at the onset of collapse, to see if asymmetries will grow from a nearly 
axisymmetric bounce. I will refer to this simulation as model RMR. 
A second motivation for taking another look at the RMR initial data comes from the 
recent results in Reference \cite{dbrown:modeone}, which suggest that a fluid body with 
toroidal density maximum (as occurs for model RMR both before and after collapse) can be
dynamically unstable to an $m=1$ mode for $T/|W|$ as low as $\sim 0.14$. The results of my 
simulation show 
that, as expected, the inner core is nearly axisymmetric immediately after core bounce. It remains 
axisymmetric until about $45\,{\rm ms}$, which is the time at which RMR stopped their simulations. 
The dominant unstable mode that begins to grow at that time is the $m=2$ bar mode, not the $m=1$ mode. 
This occurs in spite of the fact that the stability parameter has a value of around $0.19$. 

Before presenting the detailed results of the 3-D simulations, I need to establish 
some notation. The shape of the core can be described by expanding the matter density $\rho$
in spherical harmonics: 
\begin{equation}
   \rho(t,r,\theta,\phi) = \sum_{\ell = 0}^{\infty} \sum_{m=-\ell}^{m=\ell} 
   A_{\ell m}(t,r) Y_{\ell m}(\theta,\phi) \ .
\end{equation}
The quadrupole formula relates the gravitational--wave amplitude to the 
second time derivative of the quadrupole moment of the mass distribution 
\cite{dbrown:MTW}. Inserting the expansion (2) into the quadrupole formula, we 
find 
\begin{eqnarray}
   \frac{c^4 R}{2G} h^{TT}_{+} & =  &\sqrt{\frac{\pi}{5}} \sin^2\Theta \langle{\ddot A}_{20}\rangle 
  + \sqrt{\frac{2\pi}{5}} (1 + \cos^2\Theta)\, \Re( \langle{\ddot A}_{22}\rangle e^{2i\Phi} ) \cr
  & & \qquad\qquad\qquad\quad
      + \sqrt{\frac{8\pi}{15}} \sin\Theta\cos\Theta \,\Re( \langle{\ddot A}_{21}\rangle e^{i\Phi} )
   \ ,&(3a)\cr
   \frac{c^4 R}{2G} h^{TT}_{\times} & =  & -\sqrt{\frac{8\pi}{15}} 
   \cos\Theta\, \Im( \langle{\ddot A}_{22}\rangle e^{2i\Phi} ) 
   -\sqrt{\frac{8\pi}{15}} \sin\Theta\, \Im( \langle{\ddot A}_{21}\rangle e^{i\Phi} ) 
   \ ,\eqnum{3b}
\end{eqnarray}
\addtocounter{equation}{1}
for the $+$ and $\times$ components of the gravitational--wave amplitude in the transverse--traceless 
($TT$) gauge. In these formulas $R$, $\Theta$, and $\Phi$ specify the distance and angular direction 
from the source to the observation point and $\Re$ and $\Im$ denote real and imaginary parts. 
The angle brackets that appear in equations (3) are defined by 
$\langle {\ddot A}_{\ell m} \rangle = \int dr\, r^4 {\ddot A}_{\ell m}$, 
which is the spatial average of the second time derivative of $A_{\ell m}$, weighted with $r^2$. 

From equations (3) we see that only the $\ell = 2$ spherical harmonics contribute to the 
gravitational--wave signal in the quadrupole approximation. The coefficient $A_{20}$ determines the 
oblateness of the mass distribution. Note that the space average 
$\langle{\ddot A}_{20}\rangle$ appears in $h^{TT}_{+}$, but not in $h^{TT}_{\times}$. 
The coefficient $A_{22}$ corresponds 
to a bar--shaped deformation in the equatorial plane. Growth 
of this coefficient implies growth of the usual Fourier $m=2$ bar mode. 
The coefficient $A_{21}$ describes a bar--shaped deformation that is tilted out of the 
equatorial plane. This coefficient remains zero (apart from numerical noise) throughout 
the simulations, as one might expect from symmetry considerations. Note that the coefficient 
$A_{11}$, which corresponds to the $m=1$ mode 
of Reference \cite{dbrown:modeone}, does not appear in the approximate formulas (3) for 
the gravitational--wave amplitude. 

Figures 3 and 4 show the ratios $A_{20}/\rho_0$ and $|A_{22}|/\rho_0$ for model $\Omega 24$, at 
various radii. 
Here, $\rho_0$ is the average density at the given radius. For clarity of presentation, the 
results for radius $r = 60\,{\rm km}$ are shown with heavy curves. From Figure 3 we see that, 
initially, 
\begin{figure}[htb!] 
\begingroup%
  \makeatletter%
  \newcommand{\GNUPLOTspecial}{%
    \@sanitize\catcode`\%=14\relax\special}%
  \setlength{\unitlength}{0.1bp}%
\begin{picture}(3600,2160)(0,0)%
\special{psfile=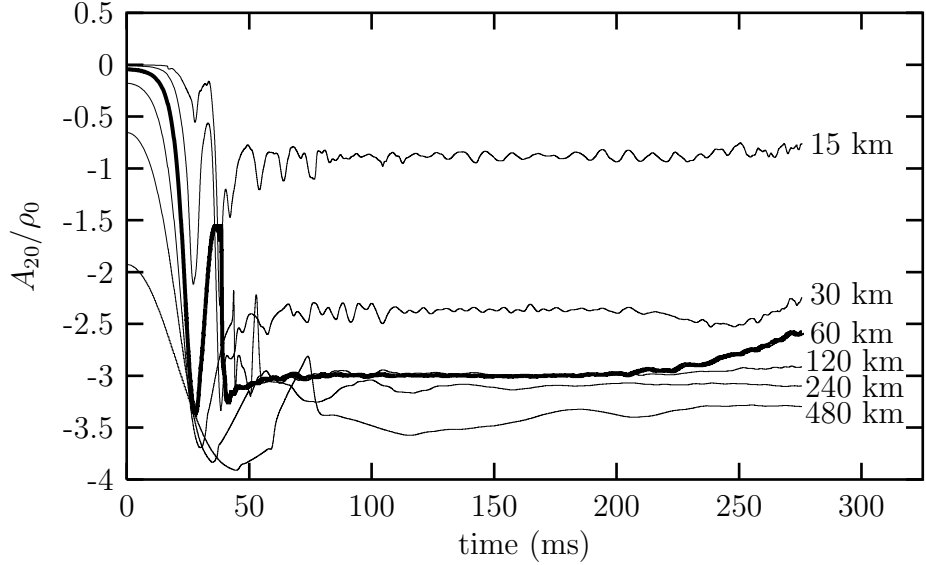 llx=0 lly=0 urx=720 ury=504 rwi=7200}
\put(3007,554){\makebox(0,0)[l]{480 km}}%
\put(3007,652){\makebox(0,0)[l]{240 km}}%
\put(3007,750){\makebox(0,0)[l]{120 km}}%
\put(3025,855){\makebox(0,0)[l]{60 km}}%
\put(3025,1004){\makebox(0,0)[l]{30 km}}%
\put(3025,1571){\makebox(0,0)[l]{15 km}}%
\put(1950,50){\makebox(0,0){time (ms)}}%
\put(100,1180){%
\special{ps: gsave currentpoint currentpoint translate
270 rotate neg exch neg exch translate}%
\makebox(0,0)[b]{\shortstack{$A_{20}/\rho_0$}}%
\special{ps: currentpoint grestore moveto}%
}%
\put(3219,200){\makebox(0,0){300}}%
\put(2758,200){\makebox(0,0){250}}%
\put(2296,200){\makebox(0,0){200}}%
\put(1835,200){\makebox(0,0){150}}%
\put(1373,200){\makebox(0,0){100}}%
\put(912,200){\makebox(0,0){50}}%
\put(450,200){\makebox(0,0){0}}%
\put(400,2060){\makebox(0,0)[r]{0.5}}%
\put(400,1864){\makebox(0,0)[r]{0}}%
\put(400,1669){\makebox(0,0)[r]{-0.5}}%
\put(400,1473){\makebox(0,0)[r]{-1}}%
\put(400,1278){\makebox(0,0)[r]{-1.5}}%
\put(400,1082){\makebox(0,0)[r]{-2}}%
\put(400,887){\makebox(0,0)[r]{-2.5}}%
\put(400,691){\makebox(0,0)[r]{-3}}%
\put(400,496){\makebox(0,0)[r]{-3.5}}%
\put(400,300){\makebox(0,0)[r]{-4}}%
\end{picture}%
\endgroup
 
\smallskip
\caption{$A_{20}/\rho_0$ vs.~$t$ for model $\Omega 24$.}
\end{figure}
\begin{figure}[htb!] 
\begingroup%
  \makeatletter%
  \newcommand{\GNUPLOTspecial}{%
    \@sanitize\catcode`\%=14\relax\special}%
  \setlength{\unitlength}{0.1bp}%
\begin{picture}(3600,2160)(0,0)%
\special{psfile=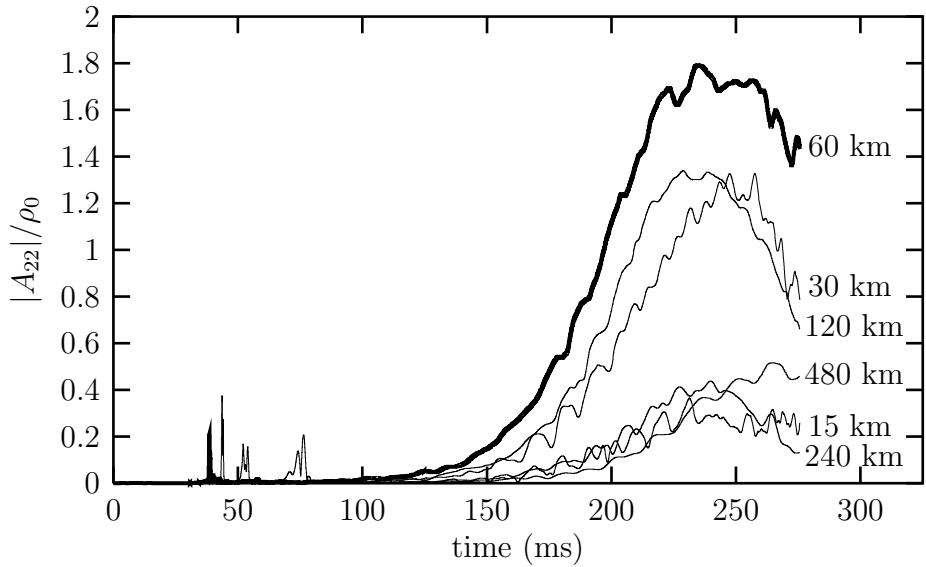 llx=0 lly=0 urx=720 ury=504 rwi=7200}
\put(3005,714){\makebox(0,0)[l]{480 km}}%
\put(3005,406){\makebox(0,0)[l]{240 km}}%
\put(3005,898){\makebox(0,0)[l]{120 km}}%
\put(3018,1576){\makebox(0,0)[l]{60 km}}%
\put(3018,1048){\makebox(0,0)[l]{30 km}}%
\put(3018,520){\makebox(0,0)[l]{15 km}}%
\put(1925,50){\makebox(0,0){time (ms)}}%
\put(100,1180){%
\special{ps: gsave currentpoint currentpoint translate
270 rotate neg exch neg exch translate}%
\makebox(0,0)[b]{\shortstack{$|A_{22}|/\rho_0$}}%
\special{ps: currentpoint grestore moveto}%
}%
\put(3215,200){\makebox(0,0){300}}%
\put(2746,200){\makebox(0,0){250}}%
\put(2277,200){\makebox(0,0){200}}%
\put(1808,200){\makebox(0,0){150}}%
\put(1338,200){\makebox(0,0){100}}%
\put(869,200){\makebox(0,0){50}}%
\put(400,200){\makebox(0,0){0}}%
\put(350,2060){\makebox(0,0)[r]{2}}%
\put(350,1884){\makebox(0,0)[r]{1.8}}%
\put(350,1708){\makebox(0,0)[r]{1.6}}%
\put(350,1532){\makebox(0,0)[r]{1.4}}%
\put(350,1356){\makebox(0,0)[r]{1.2}}%
\put(350,1180){\makebox(0,0)[r]{1}}%
\put(350,1004){\makebox(0,0)[r]{0.8}}%
\put(350,828){\makebox(0,0)[r]{0.6}}%
\put(350,652){\makebox(0,0)[r]{0.4}}%
\put(350,476){\makebox(0,0)[r]{0.2}}%
\put(350,300){\makebox(0,0)[r]{0}}%
\end{picture}%
\endgroup
 
\smallskip
\caption{$|A_{22}|/\rho_0$ vs.~$t$ for model $\Omega 24$.}
\end{figure}
the oblateness of the core increases rapidly as matter rushes inward from all directions, 
but most rapidly along the polar regions. 
($A_{20}$ is negative for an oblate spheroid, positive 
for a prolate spheroid.) At $\sim 28\,{\rm ms}$ 
the core experiences a ``polar bounce" in which the matter 
in the polar regions is reflected off the inner core, reverses direction, and forms 
shock waves that propagate outward away from the equatorial plane. During the next $\sim 10\,{\rm ms}$, 
the oblateness of the inner core decreases as matter continues to rush in from the equator and 
out from the poles. At about $38\,{\rm ms}$ the core experiences an 
``equatorial bounce" in which the matter in the equatorial plane is reflected and forms an 
outwardly propagating shock wave. As the inner core relaxes, it spreads in the equatorial 
direction and again assumes a highly oblate shape. The oblateness remains fairly constant until 
near the end of the simulation, when the bar deformation becomes strong. Figure 4 shows  
growth of the bar mode, which begins around $100\,{\rm ms}$. The spikes between $40\,{\rm ms}$ 
and $80\,{\rm ms}$ are caused by shock waves passing through the various radii. Since the shocks 
are not perfectly axisymmetric they produce relatively large but short--lived distortions 
that show up in the $A_{22}$ coefficient. The growth rate of the bar mode for model $\Omega24$ is 
$d\ln(|A_{22}|/\rho_0)/dt \approx 46/{\rm s}$. 

A graph of the ratio $|A_{22}|/\rho_0$ for model RMR shows that the bar mode begins to 
grow at about $45\,{\rm ms}$, at a rate of $d\ln(|A_{22}|/\rho_0)/dt \approx 180/{\rm s}$. 
$|A_{22}|/\rho_0$ reaches a peak value of $1.6$ at $\sim 70\,{\rm ms}$. 
The strongest bar deformation occurs within a radius of $\sim 100\,{\rm km}$. 

The graphs in Figures 5 and 6 show the gravitational--wave signals for models $\Omega 24$ and 
RMR, respectively. 
\begin{figure}[b] 
\begingroup%
  \makeatletter%
  \newcommand{\GNUPLOTspecial}{%
    \@sanitize\catcode`\%=14\relax\special}%
  \setlength{\unitlength}{0.1bp}%
\begin{picture}(3600,2160)(0,0)%
\special{psfile=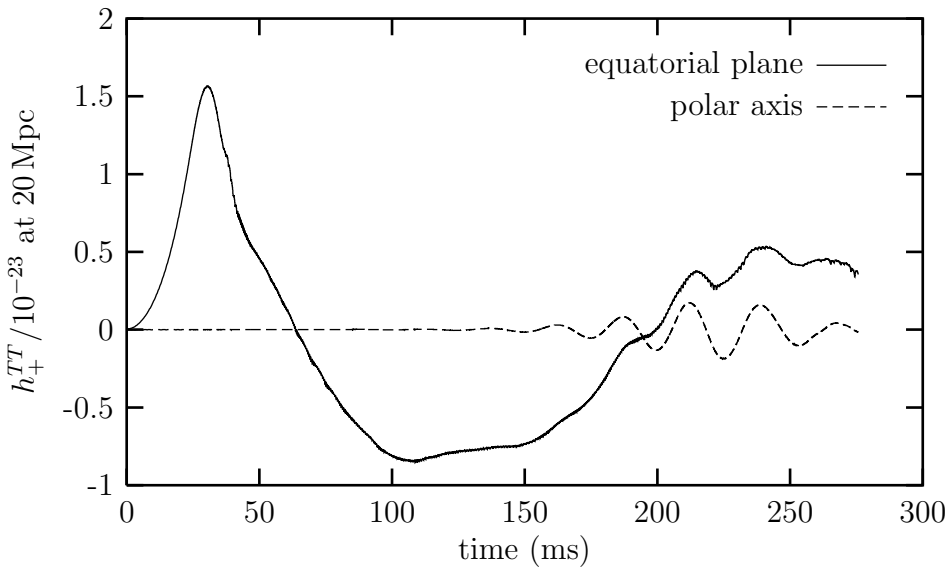 llx=0 lly=0 urx=720 ury=504 rwi=7200}
\put(3000,1727){\makebox(0,0)[r]{polar axis}}%
\put(3000,1884){\makebox(0,0)[r]{equatorial plane}}%
\put(1950,50){\makebox(0,0){time (ms)}}%
\put(100,1180){%
\special{ps: gsave currentpoint currentpoint translate
270 rotate neg exch neg exch translate}%
\makebox(0,0)[b]{\shortstack{$h^{TT}_+/10^{-23}$ at $20\,{\rm Mpc}$}}%
\special{ps: currentpoint grestore moveto}%
}%
\put(3450,200){\makebox(0,0){300}}%
\put(2950,200){\makebox(0,0){250}}%
\put(2450,200){\makebox(0,0){200}}%
\put(1950,200){\makebox(0,0){150}}%
\put(1450,200){\makebox(0,0){100}}%
\put(950,200){\makebox(0,0){50}}%
\put(450,200){\makebox(0,0){0}}%
\put(400,2060){\makebox(0,0)[r]{2}}%
\put(400,1767){\makebox(0,0)[r]{1.5}}%
\put(400,1473){\makebox(0,0)[r]{1}}%
\put(400,1180){\makebox(0,0)[r]{0.5}}%
\put(400,887){\makebox(0,0)[r]{0}}%
\put(400,593){\makebox(0,0)[r]{-0.5}}%
\put(400,300){\makebox(0,0)[r]{-1}}%
\end{picture}%
\endgroup
 
\smallskip
\caption{The gravitational wave amplitude for model $\Omega 24$.}
\end{figure}
\begin{figure}[t] 
\begingroup%
  \makeatletter%
  \newcommand{\GNUPLOTspecial}{%
    \@sanitize\catcode`\%=14\relax\special}%
  \setlength{\unitlength}{0.1bp}%
\begin{picture}(3600,2160)(0,0)%
\special{psfile=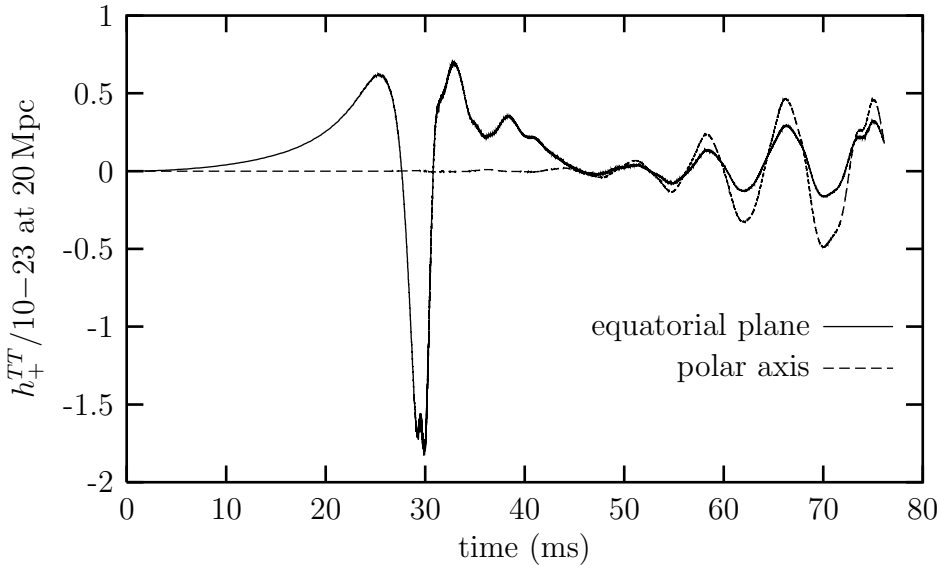 llx=0 lly=0 urx=720 ury=504 rwi=7200}
\put(3025,730){\makebox(0,0)[r]{polar axis}}%
\put(3025,887){\makebox(0,0)[r]{equatorial plane}}%
\put(1950,50){\makebox(0,0){time (ms)}}%
\put(100,1180){%
\special{ps: gsave currentpoint currentpoint translate
270 rotate neg exch neg exch translate}%
\makebox(0,0)[b]{\shortstack{$h^{TT}_+/10{-23}$ at $20\,{\rm Mpc}$}}%
\special{ps: currentpoint grestore moveto}%
}%
\put(3450,200){\makebox(0,0){80}}%
\put(3075,200){\makebox(0,0){70}}%
\put(2700,200){\makebox(0,0){60}}%
\put(2325,200){\makebox(0,0){50}}%
\put(1950,200){\makebox(0,0){40}}%
\put(1575,200){\makebox(0,0){30}}%
\put(1200,200){\makebox(0,0){20}}%
\put(825,200){\makebox(0,0){10}}%
\put(450,200){\makebox(0,0){0}}%
\put(400,2060){\makebox(0,0)[r]{1}}%
\put(400,1767){\makebox(0,0)[r]{0.5}}%
\put(400,1473){\makebox(0,0)[r]{0}}%
\put(400,1180){\makebox(0,0)[r]{-0.5}}%
\put(400,887){\makebox(0,0)[r]{-1}}%
\put(400,593){\makebox(0,0)[r]{-1.5}}%
\put(400,300){\makebox(0,0)[r]{-2}}%
\end{picture}%
\endgroup
 
\smallskip
\caption{The gravitational wave amplitude for model RMR.}
\end{figure}
The solid curves show the $+$ polarization amplitude (3a) as 
measured in the equatorial plane $\Theta = \pi/2$ at a distance of $R = 20\,{\rm Mpc}$ (the 
approximate distance to the Virgo cluster). The dashed curves show the $+$ polarization 
amplitude (3a) as measured along the rotation axis $\Theta = 0$ at $R = 20\,{\rm Mpc}$. 
Since the coefficient $A_{21}$ remains essentially zero, the $\times$ polarization amplitude 
(3b) at any angle is proportional to the $+$ amplitude at $\Theta = 0$. Observe that for both 
models, the 
gravitational waves produced by the bar---the wiggles on the graphs at late times---are 
relatively small in amplitude compared to the gravitational waves produced by the core's 
(nearly axisymmetric) collapse and bounce. 
Fourier analysis of the gravitational--wave signal for model $\Omega 24$ shows
two peaks, one around $5\,{\rm Hz}$ due to the collapse 
and bounce motion of the core, and the other around $40\,{\rm Hz}$  
due to the rotation of the bar--shaped inner core. The total energy radiated in gravitational 
waves for model $\Omega 24$, up to the end of the simulation, is only a 
${\rm few}\times 10^{-9}\,M_\odot c^2$. For model RMR, the gravitational--wave signal is spread 
across the frequency range $25$--$250\,{\rm Hz}$. The total energy radiated for this model, 
up to the end of the simulation, is a ${\rm few}\times 10^{-7}\,M_\odot c^2$. 

\section*{Discussion}

After collapse, models $\Omega 24$ and RMR are unstable to growth of the bar mode. 
On the other hand, model $\Omega 20$ is stable. The 3-D simulation of model $\Omega 20$ was 
carried out to $\sim 200\,{\rm ms}$ beyond core bounce, and the coefficient 
$A_{22}$ showed no signs of growth. At first sight these results might seem surprising:  
As shown in Figure 1, the stability parameter for
models $\Omega 24$ and $\Omega 20$ after core bounce 
exceed the nominal threshold $\sim 0.27$ 
while the stability parameter for model RMR has a sustained, post--bounce value of less than $0.20$.  
Of course, our understanding of the bar mode instability is based primarily on studies of 
isolated, equilibrium polytropes with Maclaurin--like rotation laws. There is little reason 
to believe that such a body would be a good approximation to a post--collapse stellar core 
and, indeed, for the cases studied here, it is not. The value of the stability parameter for 
the post--collapse core is not a good diagnostic for the 
presence of the bar instability. 

Inspection of the data for models $\Omega 24$ and RMR shows that only the most dense 
regions of the post--collapse core participate in growth of the bar mode. 
For the model $\Omega 24$ in particular, the bar deformation is contained within the region 
with density $\rho \gtrsim 10^{10}\,{\rm g}/{\rm cm}^3$. 
\begin{figure}[tb!] 
\epsffile{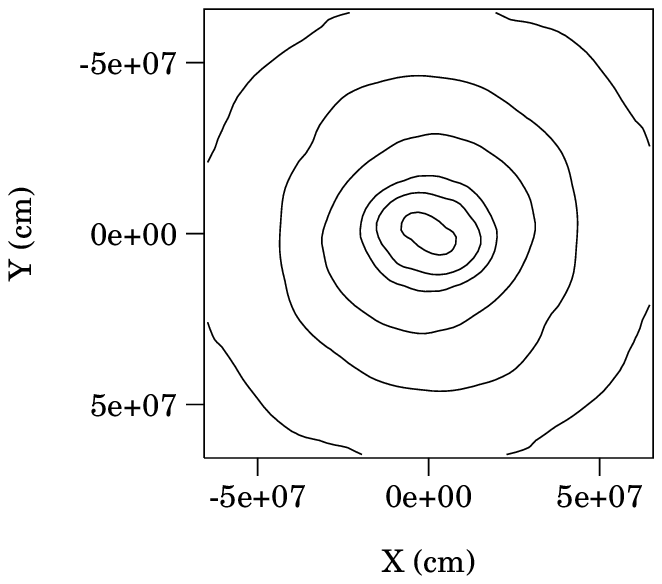}\epsffile{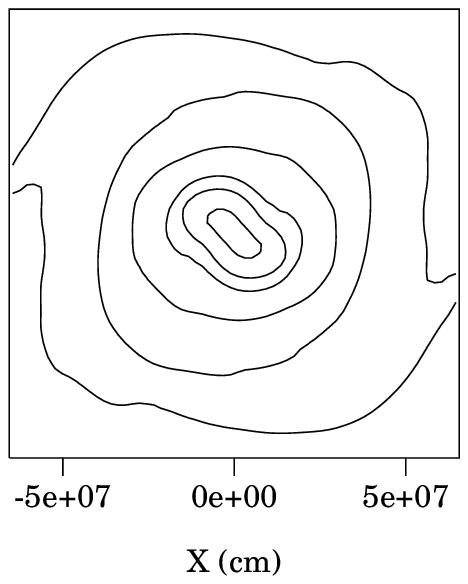}
\smallskip
\caption{Density contours for model $\Omega 24$
at $199\,{\rm ms}$ (left) and $229\,{\rm ms}$ (right). The contour 
levels, in ${\rm g}/{\rm cm}^3$, are $5.0\times 10^9$, $1.0\times 10^{10}$, 
$2.0\times 10^{10}$, $5.0\times 10^{10}$, $1.0\times 10^{11}$, and 
$5.0\times 10^{11}$.}
\end{figure}
Figure 7 shows contour 
plots of the density in the equatorial plane at $199\,{\rm ms}$, when the bar has moderate 
strength, and at $229\,{\rm ms}$, when the bar is near full strength. In the first  
plot, the contours below $\sim 10^{10}\,{\rm g}/{\rm cm}^3$ are nearly circular apart from 
some $m=4$ noise caused by the Cartesian grid. By the time of the second plot, 
the bar--shaped region with $\rho \gtrsim 10^{10}\,{\rm g}/{\rm cm}^3$ has created 
``wakes" in the surrounding matter. These wakes form 
spiral arms that trail from the ends of the bar, and give rise to the spikes seen in 
the contour at $5.0\times 10^{9}\,{\rm g}/{\rm cm}^3$. 
\begin{figure}[tb!] 
\begingroup%
  \makeatletter%
  \newcommand{\GNUPLOTspecial}{%
    \@sanitize\catcode`\%=14\relax\special}%
  \setlength{\unitlength}{0.1bp}%
\begin{picture}(3600,2160)(0,0)%
\special{psfile=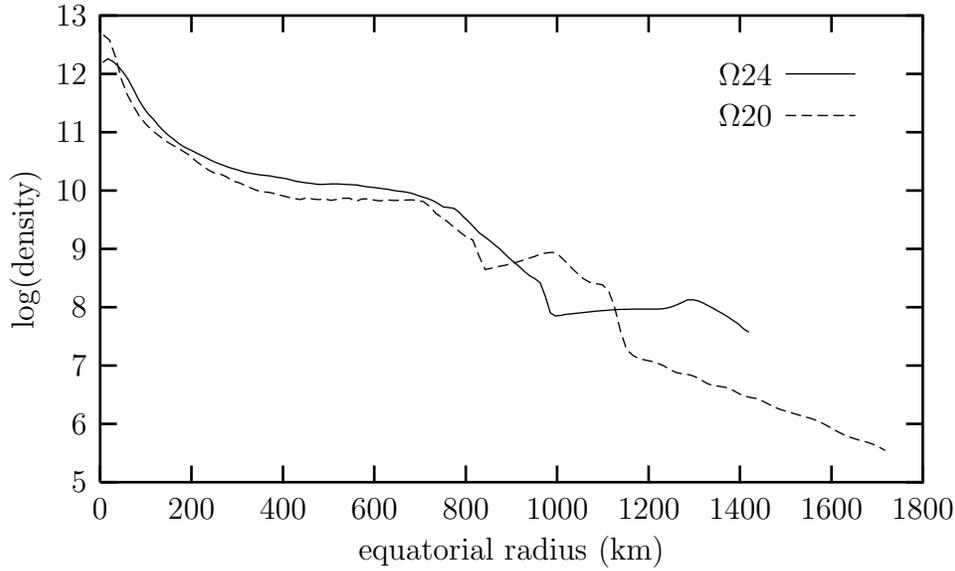 llx=0 lly=0 urx=720 ury=504 rwi=7200}
\put(2883,1683){\makebox(0,0)[r]{$\Omega 20$}}%
\put(2883,1840){\makebox(0,0)[r]{$\Omega 24$}}%
\put(1900,40){\makebox(0,0){equatorial radius (km)}}%
\put(100,1180){%
\special{ps: gsave currentpoint currentpoint translate
270 rotate neg exch neg exch translate}%
\makebox(0,0)[b]{\shortstack{log(density)}}%
\special{ps: currentpoint grestore moveto}%
}%
\put(3450,200){\makebox(0,0){1800}}%
\put(3106,200){\makebox(0,0){1600}}%
\put(2761,200){\makebox(0,0){1400}}%
\put(2417,200){\makebox(0,0){1200}}%
\put(2072,200){\makebox(0,0){1000}}%
\put(1728,200){\makebox(0,0){800}}%
\put(1383,200){\makebox(0,0){600}}%
\put(1039,200){\makebox(0,0){400}}%
\put(694,200){\makebox(0,0){200}}%
\put(350,200){\makebox(0,0){0}}%
\put(300,2060){\makebox(0,0)[r]{13}}%
\put(300,1840){\makebox(0,0)[r]{12}}%
\put(300,1620){\makebox(0,0)[r]{11}}%
\put(300,1400){\makebox(0,0)[r]{10}}%
\put(300,1180){\makebox(0,0)[r]{9}}%
\put(300,960){\makebox(0,0)[r]{8}}%
\put(300,740){\makebox(0,0)[r]{7}}%
\put(300,520){\makebox(0,0)[r]{6}}%
\put(300,300){\makebox(0,0)[r]{5}}%
\end{picture}%
\endgroup
 
\smallskip
\caption{Density profile in the equatorial plane for model 
$\Omega 24$ at $161\,{\rm ms}$ and for model 
$\Omega 20$ at $182\,{\rm ms}$.}
\end{figure}
Figure 8 shows the density as a function of radius in the equatorial plane for models 
$\Omega 20$ and $\Omega 24$. For model $\Omega 24$ the density data is taken at 
$161\,{\rm ms}$, near the beginning of the bar mode growth. 
Note that for both models, the density has a  peak in the center then levels off 
to a value of about $10^{10}\,{\rm g}/{\rm cm}^3$. Beyond $\sim 700\,{\rm km}$, the 
density drops sharply. 

Motivated by the observations above, I will define the inner core for models 
$\Omega 20$ and $\Omega 24$ to be the region interior to $\sim 10^{10}\,{\rm g}/{\rm cm}^3$.
Thus, we can view the post--collapse configuration as consisting of a dense inner core 
with $\rho \gtrsim 10^{10}\,{\rm g}/{\rm cm}^3$ surrounded 
by relatively low density material. It is this inner core 
region that is unstable for model $\Omega 24$ and stable for model $\Omega 20$. 
The inner core for model RMR can be defined roughly by 
$\rho \gtrsim 10^{10}\,{\rm g}/{\rm cm}^3$ as well. 

\begin{figure}[tb!] 
\begingroup%
  \makeatletter%
  \newcommand{\GNUPLOTspecial}{%
    \@sanitize\catcode`\%=14\relax\special}%
  \setlength{\unitlength}{0.1bp}%
\begin{picture}(3600,2160)(0,0)%
\special{psfile=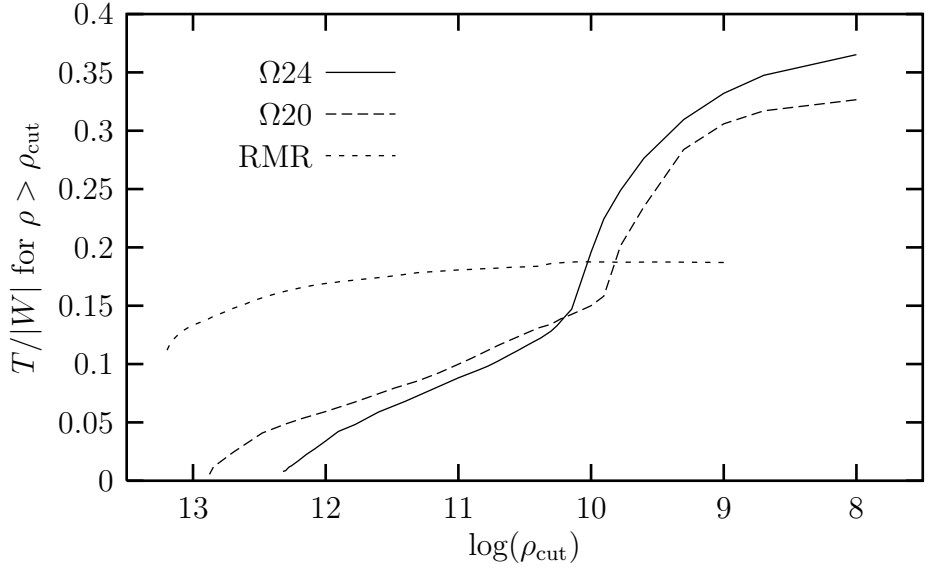 llx=0 lly=0 urx=720 ury=504 rwi=7200}
\put(1150,1526){\makebox(0,0)[r]{RMR}}%
\put(1150,1683){\makebox(0,0)[r]{$\Omega 20$}}%
\put(1150,1840){\makebox(0,0)[r]{$\Omega 24$}}%
\put(1950,50){\makebox(0,0){log($\rho_{\rm cut}$)}}%
\put(100,1180){%
\special{ps: gsave currentpoint currentpoint translate
270 rotate neg exch neg exch translate}%
\makebox(0,0)[b]{\shortstack{$T/|W|$ for $\rho > \rho_{\rm cut}$}}%
\special{ps: currentpoint grestore moveto}%
}%
\put(700,200){\makebox(0,0){13}}%
\put(1200,200){\makebox(0,0){12}}%
\put(1700,200){\makebox(0,0){11}}%
\put(2200,200){\makebox(0,0){10}}%
\put(2700,200){\makebox(0,0){9}}%
\put(3200,200){\makebox(0,0){8}}%
\put(400,2060){\makebox(0,0)[r]{0.4}}%
\put(400,1840){\makebox(0,0)[r]{0.35}}%
\put(400,1620){\makebox(0,0)[r]{0.3}}%
\put(400,1400){\makebox(0,0)[r]{0.25}}%
\put(400,1180){\makebox(0,0)[r]{0.2}}%
\put(400,960){\makebox(0,0)[r]{0.15}}%
\put(400,740){\makebox(0,0)[r]{0.1}}%
\put(400,520){\makebox(0,0)[r]{0.05}}%
\put(400,300){\makebox(0,0)[r]{0}}%
\end{picture}%
\endgroup
 
\smallskip
\caption{The stability parameter as a function of density cut--off for 
models $\Omega 24$ at $161\,{\rm ms}$, 
$\Omega 20$ at $182\,{\rm ms}$, and RMR at 
$52.9\,{\rm ms}$.}
\end{figure}
Some insights into the behavior of the three models $\Omega 20$, $\Omega 24$, and RMR 
can be gained by defining a stability parameter for the inner core, $T_{ic}/|W_{ic}|$. 
The rotational kinetic energy of the inner core $T_{ic}$ is straightforward to compute. 
The gravitational potential energy $W_{ic}$ must include the binding energy 
of the inner core material with itself, as well as the binding energy between the inner 
core material and the outer core material. 
The graph in Figure 9 shows the stability parameter of the region with density 
$\rho > \rho_{\rm cut}$, as a function of the cut--off value $\rho_{\rm cut}$, for 
the three models. The stability parameter for the inner core is obtained by setting 
$\rho_{\rm cut} \approx 10^{10}\,{\rm g}/{\rm cm}^3$. The graph shows that for model 
$\Omega 20$, $T_{ic}/|W_{ic}| \approx 0.15$ while for models $\Omega 24$ and RMR, 
$T_{ic}/|W_{ic}| \approx 0.19$. Based on these results, we might expect that model RMR is 
less stable than model $\Omega 20$. Indeed, as the 3-D simulations show, RMR is 
unstable and $\Omega 20$ is stable. Perhaps the more interesting question is this:
Why are the inner cores for $\Omega 24$ and RMR unstable, given the fact that 
the values for their stability parameters are well below $0.27$? 
The explanation might simply be that the post--collapse inner cores are not 
equilibrium polytropes with Maclaurin--like rotation laws, so the threshold for dynamical 
instability might be far different from $0.27$. Another explanation might be that a 
post--collapse inner core is not isolated, and coupling to the outer core material can drive the
bar instability. If this is correct, then growth of the bar mode is a secular process, as 
discussed for example by Schutz \cite{dbrown:Schutz}. Note that for isolated, equilibrium 
polytropes with Maclaurin--like rotation laws, the threshold for growth of the secular 
instability is about $0.14$. The inner core stability 
parameters for $\Omega 24$ and RMR are well above this threshold. On the other hand, the 
stability parameter of the inner core for model $\Omega 20$ is very close to the threshold.

\section*{Acknowledgments}

I would like to thank John Blondin for helpful discussions and for his work on 
the numerical codes. This research was supported by NSF grant PHY-0070892. Computer 
resources were provided by the North Carolina Supercomputing Center. 


\end{document}